\documentclass[11pt,oneside,onecolumn]{article}
\usepackage[]{fontenc}
\usepackage[dvips]{epsfig}
\usepackage{a4wide}

\makeatletter
\def\LyX{L\kern-.1667em\lower.25em\hbox{Y}\kern-.125emX\spacefactor1000}%
\newcommand{\lyxtitle}[1] {\thispagestyle{empty}
\global\@topnum\z@
\section*{\LARGE \centering \sffamily \bfseries \protect#1 }
}

\newcommand{\lyxletterstyle}{
\setlength\parskip{0.7em}
\setlength\parindent{0pt}
}

\makeatother

\lyxletterstyle
\begin{document}

\bfseries \large \hfill{} Teleportation and Secret Sharing with
 Pure Entangled States\hfill{} \mdseries \normalsize \bfseries \large \\
 \mdseries \normalsize 

{\hfill{}Somshubhro Bandyopadhyay\footnote{
dhom@bosemain.boseinst.ernet.in
}\hfill{} \par}

{\small \hfill{}Department of Physics, Bose Institute, 93/1 A.P.C.
Road, Calcutta -700009, India\hfill{}\hfill{}\\
\par}

{\small We present two optimal methods of teleporting an unknown qubit
using any pure entangled state. We also discuss how such methods can
also have successful application in quantum secret sharing with pure
multipartite entangled states. \par}

\( \smallskip  \)\\

\bfseries \large I. Introduction: \mdseries \normalsize \\
\\
In recent years quantum entanglement [1] has found many exciting applications
which have considerable bearing on the emerging fields of quantum
information [2] and quantum computing [3]. Two such key applications
are quantum teleportation [4] and quantum secret sharing [9]. Quantum
teleportation involves secure transfer of an unknown qubit from one
place to another and in quantum secret sharing, quantum information
encoded in a qubit is split among several parties such that only one
of them is able to recover the qubit exactly provided all the other
parties agree to cooperate. 

In quantum teleportation two parties (Alice and Bob) initially share
a maximally entangled state (for example, an EPR pair). Alice also
holds another qubit unknown to her which she wants to teleport to
Bob. For this purpose she performs certain joint two particle measurement
on her two qubits and communicates her result to Bob. Bob now applies
appropriate unitary transformations on his qubit to bring it to the
desired state. However faithful teleportation [4] (and also secure
key distribution [5]) is not possible if the entangled state used
as the quantum channel is not maximally entangled. In fact staying
within the standard teleportation scheme it is no longer possible
for Bob to reconstruct the unknown qubit exactly, with a non zero
(however small) probability. Recently, the issue of teleportation
with pure entangled states has been considered by Mor and Horodecki
[6] (originally in an earlier work of Tal Mor [7]) where they observed
that teleportation can also be understood from a more general approach
based on ``generating \( \rho  \)-ensembles at space-time separation'' by exploiting
the HJW result [8]. They introduced the concept of ``conclusive'' teleportation
and showed how perfect teleportation having a finite probability of
success is made possible with pure entangled states. By conclusive
teleportation it is meant that for certain conclusive outcomes of
some generalized measurement, perfect teleportation with fidelity
one is achieved. Of course this cannot take place with certainty unless
the state is maximally entangled. We note that the success probability
of conclusive teleportation which is twice the modulus square of the
smaller Schmidt coefficient, as obtained by Mor and Horodecki (henceforth
MH) is also optimal.

Quantum secret sharing  [9] protocol allows in splitting the quantum
information among several parties such that any one can recover the
information but not without the assistance of the remaining parties.
For simplicity all the discussions will be with three partite systems
although generalization to four or more parties is always possible.
In this case three parties (say, Alice, Bob and Charlie) initially
share a maximally entangled state, for example a GHZ state [11]. Besides
Alice also holds another qubit carrying some information (in quantum
information we know that a message is encoded in a qubit) and by performing
a Bell measurement on her two qubits she succeeds in splitting the
quantum information among Bob and Charlie. Observe that neither Bob
nor Charlie can recover the qubit in its exact form only by themselves
performing whatever local operations they wish to. Iff they agree
to act in concert, then performing certain local measurements and
communicating among themselves, any one of them can recover the desired
state. It is not possible for both to get hold of the state as it
is forbidden by the no-cloning theorem. We note that the protocol
of secret sharing is very similar to that of teleportation and in
a situation where Alice, Bob and Charlie share a non maximal entangled
state, the protocol as it is will not be successful 

In this paper we consider the issue of teleportation and secret sharing
with pure entangled state (a pure entangled state will always be taken
to be non-maximal unless stated otherwise). We suggest two more methods
for conclusive teleportation that are optimal. We refer to them as
qubit assisted conclusive teleportation process, since in both the
methods either Alice or Bob needs to prepare an ancillary qubit in
some specified state for carrying out the protocol. The motivation
behind suggesting two more methods are twofold. First one is to obtain
a possible improvement over MH suggestion from an operational point
of view with an eye towards future experiments. Secondly exploring
various explicit local strategies can also provide some insight which
can be fruitful, considering their possible application in various
other manipulations of quantum entanglement. We also show using the
methods developed for teleportation how successful secret sharing
can be implemented using pure entangled states. We will refer to this
type of secret sharing as conclusive secret sharing. 

The present paper is organized as follows. In Sec. II, we discuss
the standard teleportation scheme (BBCJPW protocol) [4] and see why
it is not successful when the shared quantum channel is a non-maximally
entangled pure state. Sec. III introduces the concept of conclusive
teleportation and the protocol of MH [6] is discussed in some detail.
In Sec. IV we present two new proposals of conclusive teleportation
and discuss relative merits of the suggested and the existing ones.
In Sec. V we describe the quantum secret sharing protocol [9]. In
Sec. VI. we discuss what we call conclusive secret sharing, i.e quantum
secret sharing with pure entangled states. There we show how the methods
developed in the preceding sections in the context of quantum teleportation
have applications in quantum secret sharing. Finally in Sec. VII we
summarize and conclude. \\

\bfseries \large II. Quantum Teleportation: BBCJPW Protocol 

\mdseries \normalsize Quantum teleportation [4] allows in sending
quantum information encoded in a qubit (a spin 1/2 particle or any
quantum two level system) from one place to another without any material
transfer of the particle itself. The two parties involved in this
process initially share a maximally entangled state. The protocol
is carried out only using local measurements (Bell measurement) and
classical communication.

Let us suppose that Alice and Bob share a maximally entangled state,
say,

\begin{equation}
\label{1}
\left| \psi \right\rangle _{AB}=\frac{1}{\sqrt{2}}\left( \left| 00\right\rangle _{AB}+\left| 11\right\rangle _{AB}\right) 
\end{equation}

and the state of the unknown qubit which Alice is supposed to send
to Bob be,

\begin{equation}
\label{2}
\left| \phi \right\rangle _{1}=a\left| 0\right\rangle +b\left| 1\right\rangle =\left( \begin{array}{c}
a\\
b
\end{array}
\right) _{1}.
\end{equation}

The combined state of the three qubits can be written as,

\begin{equation}
\label{3}
\left| \Phi \right\rangle _{1AB}=\frac{1}{2}\left( \left| \Phi ^{+}\right\rangle _{1A}\left( \begin{array}{c}
a\\
b
\end{array}
\right) _{B}+\left| \Phi ^{-}\right\rangle _{1A}\left( \begin{array}{c}
a\\
-b
\end{array}
\right) _{B}+\left| \Psi ^{+}\right\rangle _{!A}\left( \begin{array}{c}
b\\
a
\end{array}
\right) _{B}+\left| \Psi ^{-}\right\rangle _{!A}\left( \begin{array}{c}
b\\
-a
\end{array}
\right) _{B}\right) 
\end{equation}

where the states, \( \left| \Phi ^{\pm }\right\rangle ,\left| \Psi ^{\pm }\right\rangle  \) are defined by,

\[
\left| \Phi ^{\pm }\right\rangle =\frac{1}{\sqrt{2}}\left( \left| 00\right\rangle \pm \left| 11\right\rangle \right) ;\left| \Psi ^{\pm }\right\rangle =\frac{1}{\sqrt{2}}\left( \left| 01\right\rangle \pm \left| 10\right\rangle \right) \]

and form a basis (Bell-basis) in the composite Hilbert space of Alice's
two qubits.

At this stage Alice performs a measurement in the Bell basis on her
two qubits and therefore obtains any one of the four Bell states randomly
and with equal probability. She then communicates her result to Bob
(which requires 2 classical bits) who in turn rotates his qubit accordingly
to reconstruct the unknown state in its exact form. 

However, in a situation where Alice and Bob share a non-maximal but
pure entangled state of the form say,

\begin{equation}
\label{4}
\left| \psi \right\rangle _{AB}=\alpha \left| 00\right\rangle _{AB}+\beta \left| 11\right\rangle _{AB}
\end{equation}

(where, \( \alpha ^{2}+\beta ^{2}=1 \), and we assume without any loss of generality that \( \alpha ,\beta  \) to
be real with \( \alpha \geq \beta  \)) and following the standard method for teleportation,
Bob ends up with the state \( \left( \begin{array}{c}
a\alpha \\
b\beta 
\end{array}
\right)  \), which cannot be rotated back to the
desired state \( \left( \begin{array}{c}
a\\
b
\end{array}
\right)  \) without having any knowledge of the state parameters
\em a \em and \em b\em . Since the state that is teleported is supposed
to be unknown, the Bennett protocol fails to reproduce the state exactly
on Bob's side.\\

\bfseries \large III. Conclusive Teleportation: Proposal of Mor
 and Horodecki

\mdseries \normalsize Quite recently in a very interesting paper,
Mor and Horodecki [6] suggested a protocol for teleportation when
Alice and Bob share a non-maximal pure entangled state. They obtained
the optimal probability for successful teleportation which is given
by twice the modulus square of the smaller Schmidt coefficient of
the state in question. The method succeeds sometimes and when it succeeds
the fidelity is one, implying that the unknown state is exactly reproduced
on Bob's side. Following Mor and Horodecki we will continue to refer
to teleportation with pure entangled states as conclusive teleportation.

We begin with the fact that Alice and Bob share the pure entangled
state (4) and the unknown state that Alice wishes to send to Bob is
given by (2). The central feature of the scheme is to write down the
combined three qubit state in the following way,

\[
\left| \Psi \right\rangle =\left| \phi \right\rangle \left| \psi \right\rangle =\frac{1}{2}\left[ \left( \alpha \left| 00\right\rangle +\beta \left| 11\right\rangle \right) _{A}\left( \begin{array}{c}
a\\
b
\end{array}
\right) _{B}+\left( \alpha \left| 00\right\rangle -\beta \left| 11\right\rangle \right) _{A}\left( \begin{array}{c}
a\\
-b
\end{array}
\right) _{B}+\right. \]

\begin{equation}
\label{5}
\left. \left( \beta \left| 01\right\rangle +\alpha \left| 10\right\rangle \right) _{A}\left( \begin{array}{c}
b\\
a
\end{array}
\right) _{B}+\left( \beta \left| 01\right\rangle -\alpha \left| 10\right\rangle \right) _{A}\left( \begin{array}{c}
-b\\
a
\end{array}
\right) _{B}\right] 
\end{equation}

Now measurement on Alice's side takes place in two steps. The first
measurement projects the state onto either of the subspaces spanned
by \( \left\{ \left| 00\right\rangle ,\left| 11\right\rangle \right\}  \) or \( \left\{ \left| 01\right\rangle ,\left| 10\right\rangle \right\}  \). Thus this measurement has two possible outcomes that occur
with equal probability. Suppose the result is the subspace spanned
by \( \left\{ \left| 00\right\rangle ,\left| 11\right\rangle \right\}  \). Alice now performs an optimal POVM (Positive Operator Value
Measure) [10] that distinguishes conclusively between the two non-orthogonal
states \( \left( \begin{array}{c}
\alpha \\
\beta 
\end{array}
\right) _{\{00,11\}} \) and \( \left( \begin{array}{c}
\alpha \\
-\beta 
\end{array}
\right) _{\{00,11\}} \). The probability of obtaining a conclusive result is
\( 2\beta ^{2} \)(\( \beta  \) is the smaller of the Schmidt coefficients). Thus this is the
probability of successful teleportation with fidelity one. The number
of classical bits required in the above method is three. One bit is
required for Alice to inform Bob whether she is successful in discriminating
between the non orthogonal states and two more bits are required so
that Bob performs the required rotations to reconstruct the unknown
state. Note that the above proposal cannot succeed always. This is
because, there is always a possibility of an inconclusive result in
the state discrimination procedure. But when it succeeds, the probability
of success being \( 2\beta ^{2} \), the fidelity of teleportation is one. Also note
that for \( \beta =\frac{1}{\sqrt{2}} \), which corresponds to maximally entangled state, the proposal
is always successful with certainty as there need not be any inconclusive
result since in this case one discriminates between two orthogonal
states. \\

\bfseries \large IV. Qubit assisted Conclusive Teleportation \mdseries \normalsize \\
\\
We now discuss two methods for conclusive teleportation which are
optimal. We will see that both these methods can appropriately be
referred to as qubit assisted processes, since in both schemes either
Alice or Bob are required to prepare a qubit in some specified state
to implement the respective protocol. Since the process of teleportation
involves two parties, so, modifications as far as measurement and
other operations are concerned, may be suggested for any one of the
parties without introducing any new operations for the other side.
By this we mean that we can either modify the measurement part of
Alice keeping the Bob part the same i.e he only has to do the standard
rotations (proposal 1) or we can also suggest some further operations
to be carried out by Bob once the original protocol of teleportation
gets completed (proposal 2), which implies measurement part of Alice
remains unchanged. 

\bfseries IV. a. Proposal I\mdseries 

The basic idea is as follows. Alice first prepares an ancilla qubit
in a state, say \( \left| \chi \right\rangle  \) besides her usual possession of two qubits. She
now performs a certain joint three particle measurement on her three
qubits. It will be shown that for some of her results, Bob needs to
perform only the standard rotations \( (\sigma _{z},\sigma _{x},\sigma _{z}\sigma _{x}) \) to exactly reconstruct the unknown
state, after he gets some information from Alice. However, for any
of the remaining possible set of outcomes, the method works exactly
the same way as that of Mor and Horodecki, discussed in the previous
section. The method that we propose fail sometimes, but when successful
the fidelity of teleportation is one. 

Suppose, Alice and Bob share a pure entangled state given by (4) and
the state that Alice wants to teleport to Bob is given by (2). 

Alice now prepares an ancilla qubit in the state,
\begin{equation}
\label{6}
\left| \chi \right\rangle _{2}=\alpha \left| 0\right\rangle +\beta \left| 1\right\rangle =\left( \begin{array}{c}
\alpha \\
\beta 
\end{array}
\right) _{2}.
\end{equation}
Observe that the
state parameters of this ancillary qubit that Alice prepares are the
Schmidt coefficients of the pure entangled state. 

Now, the combined state of the four qubits is given by,\\

\begin{equation}
\label{7}
\left| \Psi \right\rangle _{12AB}=\left| \phi \right\rangle _{1}\otimes \left| \chi \right\rangle _{2}\otimes \left| \psi \right\rangle _{AB}=\left( \begin{array}{c}
a\\
b
\end{array}
\right) _{1}\otimes \left( \begin{array}{c}
\alpha \\
\beta 
\end{array}
\right) _{2}\otimes \left( \alpha \left| 00\right\rangle +\beta \left| 11\right\rangle \right) _{AB}
\end{equation}

and we observe that the state \( \left| \Psi \right\rangle _{12AB} \) can also be written as (we omit the
tensor product sign henceforth),

\[
\left| \Psi \right\rangle _{12AB}=\frac{\sqrt{\alpha ^{4}+\beta ^{4}}}{2}[(\alpha '\left| \Phi _{1}\right\rangle _{12A}+\beta '\left| \Phi _{2}\right\rangle _{12A})\left( \begin{array}{c}
a\\
b
\end{array}
\right) _{B}+(\alpha '\left| \Phi _{1}\right\rangle _{12A}-\beta '\left| \Phi _{2}\right\rangle _{12A})\left( \begin{array}{c}
a\\
-b
\end{array}
\right) _{B}+\]

\[
(\beta '\left| \Phi _{3}\right\rangle _{12A}+\alpha '\left| \Phi _{4}\right\rangle _{12A})\left( \begin{array}{c}
b\\
a
\end{array}
\right) _{B}+(\beta '\left| \Phi _{3}\right\rangle _{12A}-\alpha '\left| \Phi _{4}\right\rangle _{12A})\left( \begin{array}{c}
-b\\
a
\end{array}
\right) _{B}]+\]

\begin{equation}
\label{8}
\frac{\alpha \beta }{\sqrt{2}}[\left| \Phi _{5}\right\rangle _{12A}\left( \begin{array}{c}
a\\
b
\end{array}
\right) _{B}+\left| \Phi _{6}\right\rangle _{12A}\left( \begin{array}{c}
a\\
-b
\end{array}
\right) _{B}+\left| \Phi _{7}\right\rangle _{12A}\left( \begin{array}{c}
b\\
a
\end{array}
\right) _{B}+\left| \Phi _{8}\right\rangle _{12A}\left( \begin{array}{c}
-b\\
a
\end{array}
\right) _{B}]
\end{equation}

where \( \alpha '=\frac{\alpha ^{2}}{\sqrt{\alpha ^{4}+\beta ^{4}}} \) and \( \beta '=\frac{\beta ^{2}}{\sqrt{\alpha ^{4}+\beta ^{4}}} \). 

The important thing to note from Eq. (8) is that, we have succeeded
in writing down the combined state in a way such that one part clearly
resembles the one in the BBCJPW protocol (see Sec. 1) whereas the
other part resembles that of Mor and Horodeckis' (see Sec. 2). This
in turn implies that for a suitable measurement by Alice, there are
some outcomes where only standard rotations by Bob are sufficient
to construct the unknown state after he receives the result of Alice's
measurement. If this is not the case then of course one has to resort
to POVM for state discrimination. The task is now to specify the kind
of measurement that Alice should perform on her three qubits.

Observe that the following set \( \left\{ \Phi _{i}\right\} ,i=1,2...8 \), forms a complete orthonormal basis
of the combined Hilbert space of the three spin 1/2 particles (or
two level systems) that Alice holds and is defined by,

\[
\left| \Phi _{1}\right\rangle =\left| 000\right\rangle ;\left| \Phi _{2}\right\rangle =\left| 111\right\rangle ;\left| \Phi _{3}\right\rangle =\left| 011\right\rangle ;\left| \Phi _{4}\right\rangle =\left| 100\right\rangle \]
 
\begin{equation}
\label{9}
\left| \Phi _{5}\right\rangle =\frac{1}{\sqrt{2}}\left[ \left| 010\right\rangle +\left| 101\right\rangle \right] ;\left| \Phi _{6}\right\rangle =\frac{1}{\sqrt{2}}\left[ \left| 010\right\rangle -\left| 101\right\rangle \right] 
\end{equation}

\[
\left| \Phi _{7}\right\rangle =\frac{1}{\sqrt{2}}\left[ \left| 001\right\rangle +\left| 110\right\rangle \right] ;\left| \Phi _{8}\right\rangle =\frac{1}{\sqrt{2}}\left[ \left| 001\right\rangle -\left| 110\right\rangle \right] \]
\\
We now consider the following set of projection operators \( \left\{ P_{1},P_{2},P_{3},P_{4},P_{5},P_{6}\right\}  \)defined
by,

\[
P_{1}=P[\Phi _{1}]+P[\Phi _{2}];P_{2}=P[\Phi _{3}]+P[\Phi _{4}]\]

\begin{equation}
\label{10}
P_{3}=P[\Phi _{5}];P_{4}=P[\Phi _{6}];P_{5}=P[\Phi _{7}];P_{6}=P[\Phi _{8}]
\end{equation}

In principle, the measurement of an observable \( O \) is always possible
whose corresponding operator is represented by,

\begin{equation}
\label{11}
O=\sum ^{6}_{i=1}p_{i}P_{i}
\end{equation}

where Eq. (11) is the spectral decomposition of the operator \( O \). The
projectors involved in this spectral decomposition are not of same
nature. One essentially has in the set two types of projectors, both
one dimensional and two dimensional ones. \( P_{1} \) and \( P_{2} \) are the two dimensional
projectors that projects a state onto the subspaces spanned by \( \left\{ \Phi _{1},\Phi _{2}\right\}  \) and
\( \left\{ \Phi _{3},\Phi _{4}\right\}  \) respectively, whereas the rest are all one dimensional projectors.

Alice can now performs a joint three particle measurement in accordance
to Eq. (11). The possible outcomes can broadly be divided into two
types. 

\em Type a\em : If she obtains any one of the states belonging to
the set \( \left\{ \left| \Phi _{5}\right\rangle ,\left| \Phi _{6}\right\rangle ,\left| \Phi _{7}\right\rangle ,\left| \Phi _{8}\right\rangle \right\}  \), each of which occurs with probability \( \frac{\alpha ^{2}\beta ^{2}}{2} \), the state of Bob's
particle is projected onto one of the following states, \( \left( \begin{array}{c}
a\\
b
\end{array}
\right)  \), \( \left( \begin{array}{c}
a\\
-b
\end{array}
\right)  \), \( \left( \begin{array}{c}
b\\
a
\end{array}
\right)  \), \( \left( \begin{array}{c}
-b\\
a
\end{array}
\right)  \).
Qualitatively this set of outcomes resemble what we have seen in the
standard teleportation scheme. So, Alice now informs Bob the outcome
of her measurement and that requires two classical bits. Thereafter
Bob can appropriately rotate his qubit to bring it to the desired
state. 

\em Type b\em : But Alice's measurement may also project the state
onto either of the subspaces spanned by \( \left\{ \Phi _{1},\Phi _{2}\right\}  \) and \( \left\{ \Phi _{3},\Phi _{4}\right\}  \), and each such result
occurs with probability \( \frac{\left( \alpha ^{4}+\beta ^{4}\right) }{2} \). Suppose the result is the subspace spanned
by \( \left\{ \Phi _{1},\Phi _{2}\right\}  \). From (8) it follows that after such an outcome is obtained,
the combined four qubit state is given by

\begin{equation}
\label{12}
\left| \Psi \right\rangle _{12AB}=(\alpha '\left| \Phi _{1}\right\rangle _{12A}+\beta '\left| \Phi _{2}\right\rangle _{12A})\left( \begin{array}{c}
a\\
b
\end{array}
\right) _{B}+(\alpha '\left| \Phi _{1}\right\rangle _{12A}-\beta '\left| \Phi _{2}\right\rangle _{12A}])\left( \begin{array}{c}
a\\
-b
\end{array}
\right) _{B}
\end{equation}

At this stage she performs an optimal POVM measurement to conclusively
distinguish between the two states, \( \left( \begin{array}{c}
\alpha '\\
\beta '
\end{array}
\right) _{\{\Phi _{1};\Phi _{2}\}} \) and \( \left( \begin{array}{c}
\alpha '\\
-\beta '
\end{array}
\right) _{\{\Phi _{1};\Phi _{2}\}} \)(the scalar product of
these two nonorthogonal states is \( \left( \alpha '^{2}-\beta '^{2}\right)  \)). The respective positive operators
that form an optimal POVM in this subspace are:

\begin{equation}
\label{13}
A_{1}=\frac{1}{2\alpha '^{2}}\left( \begin{array}{cc}
\beta '^{2} & \alpha '\beta '\\
\alpha '\beta ' & \alpha '^{2}
\end{array}
\right) ;A_{2=}\left( \begin{array}{cc}
\beta '^{2} & -\alpha '\beta '\\
-\alpha '\beta ' & \alpha '^{2}
\end{array}
\right) ;A_{3}=\left( \begin{array}{cc}
1-\frac{\beta '^{2}}{\alpha '^{2}} & 0\\
0 & 0
\end{array}
\right) 
\end{equation}

The optimal probability of obtaining a conclusive result from such
a generalized measurement (POVM) is \( 2\beta '^{2}=\frac{2\beta ^{4}}{\alpha ^{4}+\beta ^{4}} \) . 

Suppose Alice obtains an conclusive result and therefore concludes
that the joint state of her two qubit is now \( \left( \begin{array}{c}
\alpha '\\
\beta '
\end{array}
\right) _{\{\Phi _{1};\Phi _{2}\}} \). She now informs Bob
that she had been successful in state discrimination and this requires
one classical bit. Clearly this information alone is not sufficient
for Bob because he doesn't have the information about the phase. So
Alice needs to send two more classical bits of information to enable
Bob to apply the necessary unitary transformation on his qubit. Thus
a conclusive result followed by three bits of classical information
results in perfect teleportation of the unknown qubit. 

So, given our scheme what is the probability of successful teleportation
with fidelity one? It is easy to obtain that the probability \em p
\em of having perfect teleportation is 
\begin{equation}
\label{14}
p=2\beta ^{4}+2\alpha ^{2}\beta ^{2}=2\beta ^{2}
\end{equation}

As noted earlier that this probability is the optimal probability
of perfect teleportation with a pure entangled state. We would like
to mention that number of classical bits required in this method depends
on the outcome of Alice's measurement. If her result falls in the
set when no POVM is required then number of classical bits required
is two and if it is not, the number of classical bits required is
three.

Although the above scheme may appear to be more complicated like involving
joint three particle measurement, still it simplifies the matter in
other ways. For example we have shown that there are possibilities
when no POVM is required and for those outcomes the protocol runs
exactly the same way as for a maximally entangled state. By introducing
an extra qubit this partial dependence on POVM is achieved albeit
at the cost of a joint three particle measurement. It is now clear
that an outcome falling in the set ``type a'' greatly simplifies the
remaining operations to be performed. But the probability of obtaining
an outcome of ``type a'' being \( 2\alpha ^{2}\beta ^{2} \) is always less than \( \alpha ^{4}+\beta ^{4} \), the probability
that an outcome of ``type b'' has been realized. This implies that in
more occasions Alice needs to undergo the state discrimination measurement
to achieve perfect teleportation although realization of a ``type a''
result would have simplified her task considerably . 

\bfseries IV. b. Proposal II:\mdseries 

So far we have seen that the suggested methods actually modify the
measurement part on Alice's side. But we can also think of local operations
that may be carried out by Bob after Alice performs Bell measurement
on her two qubits and communicates her result, following the standard
teleportation protocol [4]. This is what we do now. So this proposal
is carried out in two steps. In the first step the standard teleportation
scheme is followed so that the state of Bob's qubit at the end of
this, is given by \( \left( \begin{array}{c}
a\alpha \\
b\beta 
\end{array}
\right)  \). The second step involves certain local operations
to be performed by Bob. 

We first briefly discuss the CNOT operation which will be in use to
carry out the protocol. A Controlled Not gate (or quantum XOR) flips
the second spin if and only if the first spin is ``up'' i.e., it changes
the second bit iff the first bit is ``1'' \footnote{
In our notation \( \left| \uparrow \right\rangle =\left| 1\right\rangle  \) and \( \left| \downarrow \right\rangle =\left| 0\right\rangle  \).
}. It is a unitary transformation, denoted by \( U_{XOR} \), acting on pairs of
spin-1/2 and defined by the following transformation rules:

\begin{equation}
\label{15}
\left| 00\right\rangle \rightarrow \left| 00\right\rangle ;\left| 01\right\rangle \rightarrow \left| 01\right\rangle ;\left| 10\right\rangle \rightarrow \left| 11\right\rangle ;\left| 11\right\rangle \rightarrow \left| 10\right\rangle 
\end{equation}

or when written in matrix form:

\begin{equation}
\label{16}
U_{XOR}=\left( \begin{array}{cccc}
1 & 0 & 0 & 0\\
0 & 1 & 0 & 0\\
0 & 0 & 0 & 1\\
0 & 0 & 1 & 0
\end{array}
\right) 
\end{equation}

Note that CNOT gate cannot be decomposed into a tensor product of
two single bit transformation.

The method that we propose now is as follows: Recall that following
the Bennett protocol when Alice and Bob shares a pure entangled state,
Bob ends up with a state given by \( \left( \begin{array}{c}
a\alpha \\
b\beta 
\end{array}
\right)  \). Till this stage the number of
classical bits required is two and no more bits will be required because
all operations will now be carried out by Bob and there is no need
to communicate any further with Alice. 

We start from this stage when the state of Bob's qubit (we refer this
qubit as ``qubit 1'' for convenience) is given by \( \left( \begin{array}{c}
a\alpha \\
b\beta 
\end{array}
\right) _{1} \) and suggest the
following local operations. 

Bob prepares an ancilla qubit (qubit 2) in a state \( \left| 0\right\rangle _{2} \). Thus the combined
state of the two qubits that Bob holds is now given by,

\begin{equation}
\label{17}
\left| \Psi \right\rangle _{12}=a\alpha \left| 00\right\rangle _{12}+b\beta \left| 10\right\rangle _{12}
\end{equation}

Bob now performs a CNOT operation on his two qubit state, thus transforming
it into the state ,

\begin{equation}
\label{18}
\left| \Psi \right\rangle _{12}=a\alpha \left| 00\right\rangle _{12}+b\beta \left| 11\right\rangle _{12}
\end{equation}

Thus the two particles become entangled and this is absolutely necessary.
The whole idea is to entangle the particle with an ancilla and then
perform some measurement which serves the purpose. Now observe that
the state given by (18) can also be written as,

\begin{equation}
\label{19}
\left| \Psi \right\rangle _{12}=\frac{1}{2}\left[ \left( \alpha \left| 0\right\rangle +\beta \left| 1\right\rangle \right) _{1}\left( \begin{array}{c}
a\\
b
\end{array}
\right) _{2}+\left( \alpha \left| 0\right\rangle -\beta \left| 1\right\rangle \right) _{1}\left( \begin{array}{c}
a\\
-b
\end{array}
\right) _{2}\right] 
\end{equation}

From (19) it is clear that a state discrimination measurement which
can conclusively distinguish between the two non orthogonal states
\( \alpha \left| 0\right\rangle +\beta \left| 1\right\rangle  \) and \( \alpha \left| 0\right\rangle -\beta \left| 1\right\rangle  \) will give the desired result. In the last subsection we have
discussed in some detail the formalism and the respective operators
involved in such a measurement. So we don't give the explicit representation
here. Now, this optimal state discrimination measurement which is
an optimal POVM measurement can be carried out on any one of the two
qubits that Bob holds and let us assume that it is qubit 1 on which
such a measurement is performed. As we have seen earlier that the
optimal probability of a conclusive result is \( 2\beta ^{2} \). It is clear that
this is also being the probability of perfect teleportation with fidelity
one, because a conclusive outcome implies that the state of qubit
2 is now given by either \( \left( \begin{array}{c}
a\\
b
\end{array}
\right)  \) or \( \left( \begin{array}{c}
a\\
-b
\end{array}
\right)  \) depending on the state of qubit 1.
For example, suppose Bob concludes that the state of qubit 1 after
his POVM measurement is \( \left( \begin{array}{c}
\alpha \\
\beta 
\end{array}
\right)  \), then with certainty he also concludes that
the state of qubit 2 is now what he desired for. Thus this method
also produces the optimal probability of successful teleportation.
\\

{\bfseries \large V. Quantum Secret Sharing 
\par}

In quantum secret sharing [9] a person splits quantum information
(encoded in qubits) among several other persons such that no individual
can recover the whole information unless properly aided by the rest.
This is another useful application of quantum entanglement and can
play important roles in various diverse practical scenarios (see ref.
[9]). For simplicity we will be explicit only in three partite systems
but the methods can nevertheless be generalized to any number of parties.

The protocol of quantum secret sharing is as follows. Three parties,
say, Alice, Bob and Charlie initially share a maximally entangled
state, for example, a GHZ state [11],

\begin{equation}
\label{20}
\left| \psi \right\rangle _{ABC}=\frac{1}{\sqrt{2}}(\left| 000\right\rangle _{ABC}+\beta \left| 111\right\rangle _{ABC})
\end{equation}

Alice also possesses another qubit, say, \( \left( \begin{array}{c}
a\\
b
\end{array}
\right)  \). Alice performs a Bell
measurement on her two qubits and communicates her result to Bob and
Charlie who in turn can perform appropriate rotations on their respective
qubits so that the pure entangled state that they now share can be
written as,

\begin{equation}
\label{21}
\left| \Psi \right\rangle _{BC}=a\left| 00\right\rangle _{BC}+b\left| 11\right\rangle _{BC}
\end{equation}

Since information can be encoded in the state of a qubit, by performing
the Bell measurement Alice actually splits the information which is
now shared via the pure entangled state (21) between Bob and Charlie.
The important thing to note that neither Bob nor Charlie can recover
the state \( \left( \begin{array}{c}
a\\
b
\end{array}
\right)  \) by any general operations on their respective sides without
communicating among themselves. They individually do not have any
useful information whatsoever. Though they have the amplitude information
but that is not sufficient since information about the phase is not
available. So, in this situation only one of the parties (either Bob
or Charlie) will be able to reconstruct the state provided the other
party agrees to cooperate. Assuming that they do agree to work in
tandem and they also agree on the person (let us assume it is Charlie)
going to have the state, the remaining part of the protocol now goes
like this. First we rewrite the state given by (21) in the following
way,

\begin{equation}
\label{22}
\left| \Psi \right\rangle _{BC}=\frac{1}{\sqrt{2}}\left[ \frac{1}{\sqrt{2}}\left( \left| 0\right\rangle +\left| 1\right\rangle \right) _{B}\left( \begin{array}{c}
a\\
b
\end{array}
\right) _{C}+\frac{1}{\sqrt{2}}\left( \left| 0\right\rangle -\left| 1\right\rangle \right) _{B}\left( \begin{array}{c}
a\\
-b
\end{array}
\right) _{C}\right] 
\end{equation}

Bob performs a measurement on his qubit in the \( x \)-basis where the \( x- \)
eigenstates are defined by,

\begin{equation}
\label{23}
\left| x\pm \right\rangle =\frac{1}{\sqrt{2}}\left( \left| 0\right\rangle \pm \left| 1\right\rangle \right) 
\end{equation}

and communicates his outcome to Charlie. This requires only one bit
of information. Charlie now can appropriately rotate his qubit to
reconstruct the unknown state. Note that the protocol is very similar
to that of teleportation. Now, it is easy to see that, instead of
sharing a GHZ state if Alice, Bob and Charlie initially shared a non
maximally entangled state of the form,

\begin{equation}
\label{24}
\left| \psi \right\rangle _{ABC}=\alpha \left| 00\right\rangle _{ABC}+\beta \left| 11\right\rangle _{ABC}
\end{equation}

then following the protocol as it is Charlie ends up with the state
\( \left( \begin{array}{c}
a\alpha \\
b\beta 
\end{array}
\right)  \). But this is not the state that Charlie wishes to have. Recall that
we faced a similar situation in the case of quantum teleportation
and the similarity between the nature of these two processes indicate
the possibility of successful application of the methods developed
for teleportation in this scenario. Indeed we will see that the methods
discussed in the previous sections can be suitably applied so that
secret sharing becomes ultimately successful with a nonzero probability.
As we shall also see that in this case also the probability of successful
secret sharing will turn out to be \( 2\beta ^{2} \) and is conjectured to be optimal.
This is the subject of the next section.\\

{\bfseries \large VI. Quantum Secret Sharing with Pure Entangled
 States: Conclusive Quantum Secret Sharing
\par}

Note that we can broadly view the information splitting process as
a method carried out in three stages.

First stage: Measurement by Alice and communication of her outcome
to Bob and Charlie. Bob and Charlie rotates their respective qubit
so that their state is given by (21).

Second stage: Measurement by Bob and communication of his result to
Charlie.

Final stage: Charlie performs some unitary transformation on his qubit
if necessary.

Since our goal is to implement secret sharing successfully with non
maximal entangled states we can suggest modifications at any one such
stage. We propose three explicit schemes for this purpose. To be explicit,
the first scheme changes the type of measurement by Alice only, keeping
the remaining part of the original protocol intact. The second one
keeps the measurement part of Alice intact but modifies that of Bob
and the last method keeps the whole protocol intact till Charlie's
end and suggest further local operations to be carried out by him.
We won't describe the first and the last scheme in details because
the methods that have been developed (including that of Mor and Horodecki)
will be used and there is no qualitative difference with teleportation
as far as their application is concerned. The second proposal will
be described in detail. As we shall see later, we can appropriately
call such quantum secret sharing as conclusive quantum secret sharing
because the success of the protocols depend on conclusive outcomes
of some generalized measurements. 

We begin with the fact that, Alice, Bob and Charlie share a pure entangled
state of the form (24). 

\bfseries VI. a. Proposal I\mdseries 

The goal of this proposal is to modify measurement part of Alice so
that after Alice carries out her specific measurement and communicates
her result, the state of Bob and Charlie will be given by (21). If
this is achieved then the remaining part goes exactly as in the original
protocol of Hillary et. al. [9]. To achieve this purpose we note that
two methods that have already been suggested for teleportation with
pure entangled states, may turn out to be useful. Indeed, when Alice
performs a measurement on her qubits either following the MH protocol
(Sec. 2) or Proposal I of QACT (qubit assisted conclusive teleportation)
scheme, in either case after she communicates her outcome to Bob and
Charlie, the state shared by Bob and Charlie is given by (21), which
is precisely what we intended to achieve. It is easy to see that the
probability of such conclusive secret sharing is \( 2\beta ^{2} \).

\bfseries VI. b. Proposal II\mdseries 

The previous proposal suggested changes in the type of measurement
by Alice. In this proposal we keep that part same as that in [9],
i.e., Alice first performs a Bell measurement on her two qubits and
so on. Since, now Alice, Bob and Charlie initially shared a pure entangled
state (24), then after completion of the first stage of the protocol
[9], the entangled state shared by Bob and Charlie will be,

\begin{equation}
\label{25}
\left| \Psi \right\rangle _{BC}=a\alpha \left| 00\right\rangle _{BC}+b\beta \left| 11\right\rangle _{BC}
\end{equation}

instead of (21). 

This state (25) can also be written as,

\begin{equation}
\label{26}
\left| \Psi \right\rangle _{BC}=\frac{1}{2}\left[ \left( \alpha \left| 0\right\rangle +\beta \left| 1\right\rangle \right) _{B}\left( \begin{array}{c}
a\\
b
\end{array}
\right) _{C}+\left( \alpha \left| 0\right\rangle -\beta \left| 1\right\rangle \right) _{B}\left( \begin{array}{c}
a\\
-b
\end{array}
\right) _{C}\right] 
\end{equation}

Now, a conclusive result of a POVM measurement (discussed in Sec.
IV. a., for details see [10]) by Bob to discriminate between the two
non orthogonal states, \( \alpha \left| 0\right\rangle +\beta \left| 1\right\rangle  \) and \( \alpha \left| 0\right\rangle -\beta \left| 1\right\rangle  \) is sufficient. It is clear from (26)
that, when Bob concludes that the state of his qubit is \( \alpha \left| 0\right\rangle +\beta \left| 1\right\rangle  \) (or \( \alpha \left| 0\right\rangle -\beta \left| 1\right\rangle  \)),
the state of Charlie's qubit is projected onto \( a\left| 0\right\rangle +b\left| 1\right\rangle  \) (or \( a\left| 0\right\rangle -b\left| 1\right\rangle  \)). But for Charlie
to have this information Bob needs to communicate with him and he
needs to do that twice. First he informs if he is successful (requires
one classical bit) and if he is, then he notifies his result (requiring
1 classical bit) so that Charlie can perform appropriate rotation
if necessary. 

The probability of this being successful is nothing but the probability
of obtaining a conclusive result from the state discrimination measurement.
So the probability of being successful is \( 2\beta ^{2} \). 

\bfseries VI. c. Proposal III\mdseries 

This proposal follows the original secret sharing protocol to its
full so that at the end the state of Charlie's qubit is \( a\alpha \left| 0\right\rangle +b\beta \left| 1\right\rangle  \). In fact
in teleportation also when the standard scheme was followed the state
of Bob's qubit resulted in the same state. So we can successfully
apply here, the proposal II of QACT (qubit assisted conclusive teleportation)
to recover the desired state which is discussed in details in Sec.
IV. B. Again the probability of being successful is \( 2\beta ^{2} \).\\

{\bfseries \large VII. Summary and Conclusion\par}

In summary, we have described two optimal methods for teleporting
an unknown quantum state using any pure entangled state. A positive
implication of one of our strategies is in its partial dependence
on POVM to achieve perfect teleportation where we have seen that for
some of Alice's outcomes, only standard rotations are to be performed
by Bob to get the unknown state. Nevertheless the cost one has to
pay for it is a joint three particle measurement. The number of classical
bits required is three if it gets necessary to perform a POVM measurement,
otherwise it is two. The second strategy reduces the number of classical
bits to two, since the local operations are carried out by Bob after
following the standard teleportation scheme. 

We have also discussed how the methods developed for conclusive teleportation
can be successfully applied in quantum secret sharing in a situation
where the parties share a pure non maximal entangled state among themselves.
We call it as conclusive secret sharing analogous to conclusive teleportation.
Here we have exploited the qualitative similarity between the two
processes of teleportation and secret sharing. The success probability
of such conclusive secret sharing also happens to be twice the square
of the smaller Schmidt coefficient (the three partite entangled state
in consideration is Schmidt decomposable) and is conjectured to be
optimal. \\

{\bfseries \large Acknowledgments\par}

I wish to acknowledge Guruprasad Kar and Anirban Roy for many stimulating
discussions. I thank Ujjwal Sen for careful reading of the manuscript..\\

{\bfseries \large References\par}

[1] E. Schrodinger, Naturwissenschaften \bfseries 23\mdseries ,
807 (1935); \bfseries 23\mdseries , 823 (1935); \bfseries 23\mdseries ,
844 (1935); For a review see: M. B. Plenio and V. Vedral, Cont. Phys.
\bfseries 39\mdseries , 431 (1998); Lecture notes of Lucien
hardy available at http://www.qubit.org.

[2] C. H. Bennett, Physics Today \bfseries 48\mdseries , 24
(1995); J. Preskill, Proc. Roy. Soc. A: Math., Phys. and Eng. \bfseries 454\mdseries ,
469 (1998).

[3] A brief but excellent article is by R. Jozsa, quant-ph/9707034;
C. H. Bennett and D. DiVincenzo, Quantum Computing: Towards an engineering
era?, Nature, \bfseries 377\mdseries , 389 (1995).

[4] C. H. Bennett, G. Brassard, C. Crepeau, R. Jozsa, A. Peres and
W. K. Wootters, Phys. Rev. Lett. \bfseries 70\mdseries , 1895
(1993).

[5] A. K. Ekert, Phys. Rev. Lett. \bfseries 67\mdseries , 661
(1991).

[6] T. Mor and P. Horodecki, quant-ph/9906039. 

[7] T. Mor, quant-ph/ 9608005.

[8] L. P. Hughston, R. Jozsa, W. K. Wootters, Phys. Letts. A, \bfseries 183\mdseries ,
14 (1993).

[9] M. Hillery, V. Buzek and A. Berthiaume, Phys. Rev. A, \bfseries 59\mdseries ,
1829 (1999).

[10] A. Peres, Quantum Theory: Concepts and Methods (Kluwer, Dordrecht,
1993), Ch. 9.

[11] D. M. Greenberger, M. A. Horne, A. Shimony, and A. Zeilinger,
Am. J. Phys. \bfseries 58, \mdseries 1131 (1990).

\end{document}